\documentclass[10pt,conference]{IEEEtran}
\IEEEoverridecommandlockouts
\usepackage{cite}
\usepackage{amsmath,amssymb,amsfonts}
\usepackage{algorithmic}
\usepackage{graphicx}
\usepackage{textcomp}
\usepackage{xcolor}
\usepackage{booktabs}
\usepackage{multirow}
\usepackage{listings} 
\usepackage[most]{tcolorbox}
\usepackage{inconsolata}

\newtcblisting[auto counter]{sexylisting}[2][]{sharp corners, fonttitle=\bfseries, colframe=gray, listing only, listing options={linewidth=\columnwidth,breaklines=true,basicstyle=\small,language=java},  title=Listing \thetcbcounter: #2, #1}

\def\BibTeX{{\rm B\kern-.05em{\sc i\kern-.025em b}\kern-.08em
    T\kern-.1667em\lower.7ex\hbox{E}\kern-.125emX}}
\begin{document}

\title{iContractBot: A Chatbot for Smart Contracts' Specification and Code Generation\\
}
\DeclareRobustCommand*{\IEEEauthorrefmark}[1]{%
  \raisebox{0pt}[0pt][0pt]{\textsuperscript{\footnotesize\ensuremath{#1}}}}

\author{\IEEEauthorblockN{Ilham Qasse\IEEEauthorrefmark{1},
Shailesh Mishra\IEEEauthorrefmark{2}, 
Mohammad Hamdaqa\IEEEauthorrefmark{1} \,\IEEEauthorrefmark{3}
}
\IEEEauthorblockA{\IEEEauthorrefmark{1}Department of Computer Science,
Reykjavik University, Reykjavik, Iceland \\
\IEEEauthorrefmark{2}Department of Electrical Engineering, Indian Institute of Technology Kharagpur, Kharagpur, India \\ 
\IEEEauthorrefmark{3}Department of Computer and Software Engineering, Polytechnique Montreal, Montreal, Canada \\
\IEEEauthorrefmark{1}\{ilham20$,$mhamdaqa\}@ru.is, \IEEEauthorrefmark{2}mshailesh0511@iitkgp.ac.in, \IEEEauthorrefmark{3}mhamdaqa@polymtl.ca
}
}
\maketitle

\begin{abstract}
Recently, Blockchain technology adoption has expanded to many application areas due to the evolution of smart contracts. However, developing smart contracts is non-trivial and challenging due to the lack of tools and expertise in this field. A promising solution to overcome this issue is to use Model-Driven Engineering (MDE), however, using models still involves a learning curve and might not be suitable for non-technical users. To tackle this challenge, chatbot or conversational interfaces can be used to assess the non-technical users to specify a smart contract in gradual and interactive manner.

In this paper, we propose iContractBot, a chatbot for modeling and developing smart contracts. Moreover, we investigate how to integrate iContractBot with iContractML, a domain-specific modeling language for developing smart contracts, and instantiate intention models from the chatbot. The iContractBot framework provides a domain-specific language (DSL) based on the user intention and performs model-to-text transformation to generate the smart contract code. A smart contract use case is presented to demonstrate how  iContractBot can be utilized for creating models and generating the deployment artifacts for smart contracts based on a simple conversation.
\end{abstract}


\section{Introduction}\label{intro}
Smart contracts are self-executed program codes that are hosted on a blockchain platform, to enforce agreements when conditions are met \cite{kolvart2016smart}. Smart contracts are considered a great advancement for blockchain technology, as it enabled the technology to be adopted in many fields such as finance, identity management, Internet of Things, etc \cite{zheng2020overview}. However,  developing smart contract code is challenging especially for non-technical users \cite{zheng2020overview,hewa2020survey}, as it requires one to understand (i) the language used to code the smart contract, (ii) the infrastructure constraints and limitations, and (iii) the relationships between the deployed artifacts and the resources. 

Model-Driven Engineering (MDE) is one of the popular approaches used to address smart contract development challenges \cite{syahputra2019development,tran2018lorikeet,hamdaqa2020icontractml}. MDE is a software development methodology where models are used as first class entities for software development. Models are constructed representing distinct perspectives on a software system. They may be refined, developed into a new version, and can be applied to create executable code. The main goal is to elevate the extent of abstraction and to broaden and evolve complex software program structures utilizing models only.

In previous work \cite{hamdaqa2020icontractml}, we proposed iContractML, a graphical modeling framework to develop and generate smart contracts code. While graphical interaction mechanisms are famous and widely accepted, some users may lack the technical abilities required to use them \cite{perez2019towards}. Moreover, using MDE requires a steep learning curve and might be challenging for non-technical users who are not familiar with modeling tools or DSLs \cite{bucchiarone2020grand, tolvanen2016model}. Chatbot is a promising solution to tackle this issue, where it can be utilized to facilitate non-technical users to use MDE and to enhance usability and user experience \cite{valtolina2020communicability,perez2019flexible}.

In this paper, we explore the use of chatbots  to develop and model smart contracts instead of the graphical interface used in iContractML \cite{hamdaqa2020icontractml}. 
The main contribution of this paper is :
\begin{itemize}
\item Integrate the chatbot application with the model-driven based framework iContractML.
\item iContractBot: a goal-oriented chatbot application to allow users (technical or non-technical) to develop their smart contract in a gradual and interactive manner.
\end{itemize}

The rest of this paper is organized as follows: Section \ref{icontractbot-design} describes the system design of iContractBot. The implementation of iContractBot is presented in Section \ref{icontractbot-imp}. Finally, Section \ref{conclusion} concludes the paper.
\begin{figure*}
\centerline{\includegraphics[scale=0.7]{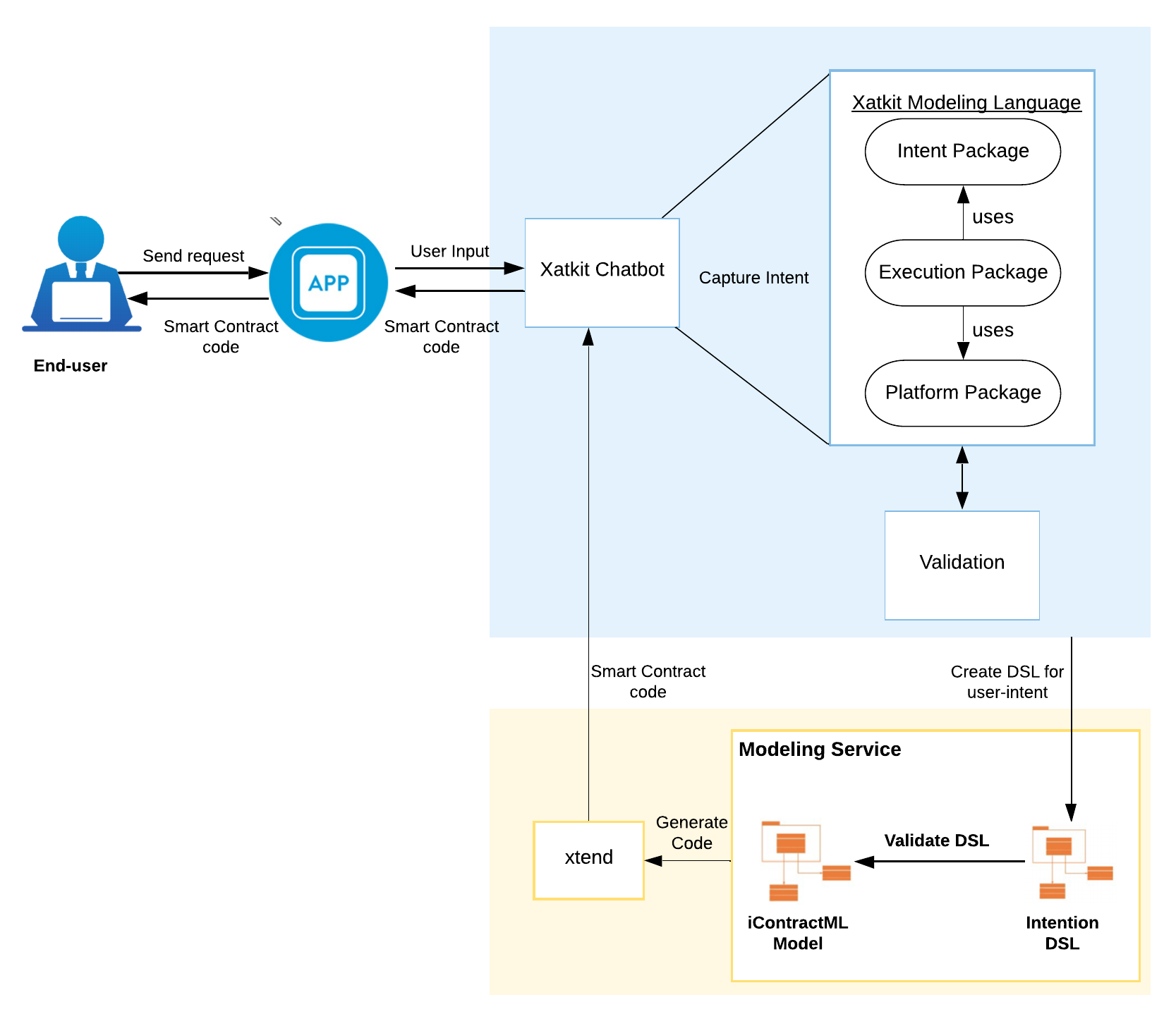}}
\caption{iContractBot system architecture}
\label{sys}
\end{figure*}
\section{iContractBot System Design}\label{icontractbot-design}
The main goal of this paper is to create a conversational agent for model specification, where we have the conversation with the smart contract developer as an input and model specification as output.  We aim to provide a link between the natural language conversation and modelling specifications, which includes capturing and extracting modeller intent, mapping the intention to modeling actions, and validating the model. 

\subsection{Modeller Intention Detection}
Any goal-oriented conversational framework requires an intent recognition component to understand the user's goal or objective. The bot must classify the end-user's utterance into one of the predefined intents. There are many chatbot frameworks available to build conversational bots and to detect intents, such as Google Dialogflow \footnote{https://dialogflow.cloud.google.com}, IBM Watson Assistant \footnote{https://www.ibm.com/cloud/watson-assistant}, etc.  Xatkit \cite{daniel2020xatkit} is an open-source framework that supports integration with the previously mentioned platforms to capture user intent and understand advanced natural language. Moreover, this framework empowers building platform-independent chatbots \cite{daniel2020xatkit}. Hence, in this paper, we have adopted the Xatkit bot framework to build the conversational bot, and to detect the user input. The user intent is detected based on a predefined set of expressions. 

\subsection{Modeller Intention to Modelling Specification Mapping}
To link natural language conversation and modeling specifications, we need to map the detected user intent to the model specification (iContractML). In order to provide this link, there is a need to identify the principal entities of the model specification and represent it as a key structure for mapping it to the captured user intent. A DSL model is created based on the mapping of the user intent and the structure of the model specification.

\subsection{Model Validation}
In MDE, data validation is important because it guarantees that the system runs on valid and meaningful data \cite{rossini2011formal}. The entire model-based development process can be faulty due to a single inaccurate input data.
In this paper, we are performing input sanitation at the chatbot level and output validation for the created DSL, before any model transformation. This facilitates the data validation process as a chatbot is an open input environment that is more flexible to validate, unlike MDE environments. We validate the detected user intention from any contextual errors or missing data based on the defined structure of the DSL model. This validation is done based on pre-defined rules that will enable the chatbot to handle incomplete or inconsistent elements (e.g., missing relationships) defined by the user. 

\section{iContractBot Implementation}\label{icontractbot-imp}
In our previous work \cite{hamdaqa2020icontractml}, we have created a unified reference model for smart contracts. Moreover, we proposed iContractML which is a graphical framework to develop smart contracts onto multiple blockchain platforms. iContractBot integrates with the reference model of iContractML to generate smart contracts code through the chatbot framework instead of the graphical interface. iContractBot integrates different tools, including Xatkit bot framework \cite{daniel2020xatkit}, Xtext \footnote{https://www.eclipse.org/Xtext/}, and Xtend \footnote{https://www.eclipse.org/xtend/}. 
Figure \ref{sys} demonstrates the main components of the iContractBot, which are :
\begin{itemize}
\item Xatkit chatbot: a chatbot framework we used to implement the conversational bot and to capture the user intent. The user intent is the smart contract description provided by the end-user. 
\item Validation entity: validates the captured user intent against a set of predefined validation rules, and notifies the end-user if any extra details are required.
\item Modeling Service: consists of an iContractML model and a generated DSL file based on the user intent. 
\item Xtend: used to generate the smart contract code based on the selected blockchain platform.
\end{itemize}

\subsection{Preliminaries and Running Example}\label{example}
Using iContractBot, we have created models and generated the deployment artifacts for a vehicle auction use case. In this use case, a smart contract is used to auction vehicles, where the vehicle is the key asset. There are two participants in this example: owner and bidder. The smart contract is created by the owner to auction his/her vehicle. The bidder can place bids on the vehicles that they are interested in. 

\subsection{Chatbot Framework}\label{framework}
\begin{figure}
\centerline{\includegraphics[scale=0.75]{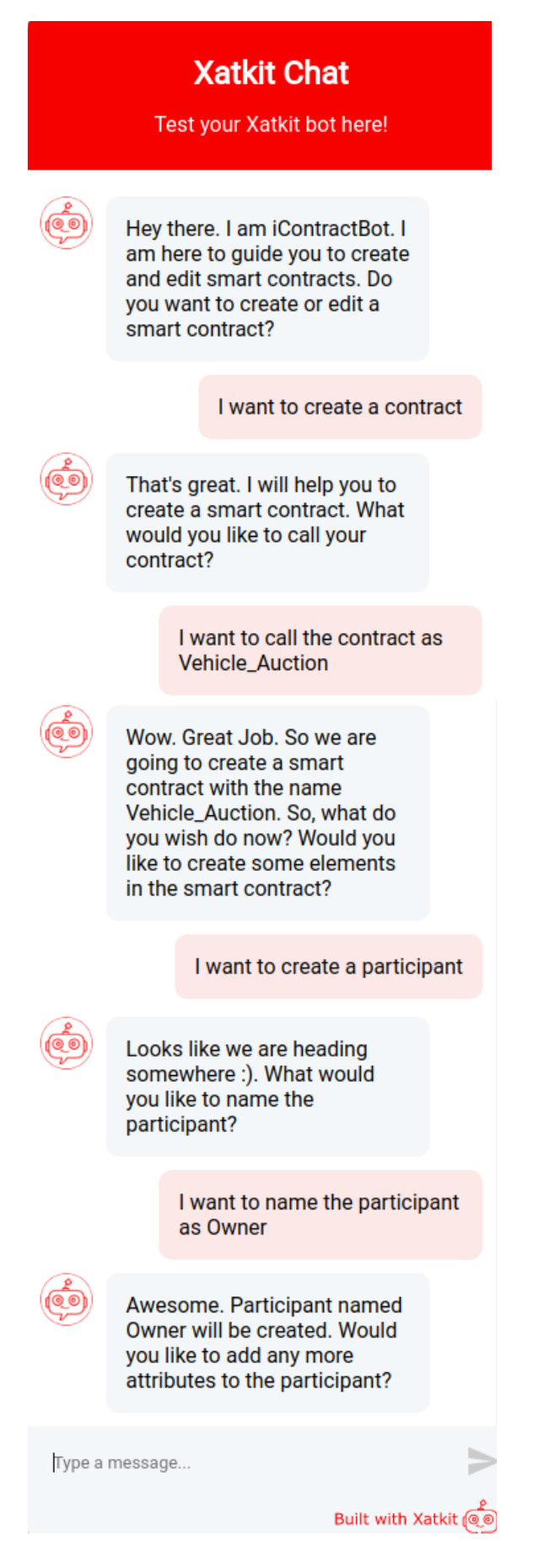}}
\caption{Vehicle Auction Use Case Example}
\label{VA}
\end{figure}

 Xatkit is an open-source framework to easily build platform-independent chatbots. We have used this framework to develop a web-based conversational bot and to capture the user intent. The user intent represents the smart contract use case that the end-user is interested to generate. The chat flow in the bot is directed based on the main components of the reference model of the smart contract \cite{hamdaqa2020icontractml}. Once the end-user finalizes the smart contract use case, iContractBot will validate the captured user intent. If there are any missing required details or errors in the user request, the bot will notify the end-user. 
A sample of the conversation to build the vehicle auction use case is illustrated in Figure \ref{VA}.

\subsection{Modeling Service}\label{modelling-service}
 The modeling service represents the linking of the user intention to the modeling specification of iContractML. We first identified the key classes of iContractML meta-model and create a DSL model (illustrated in Figure \ref{DSL}) that acts as a structure for mapping and validating. The created DSL model is as follows:
 \begin{figure*}
\centerline{\includegraphics[scale=0.69]{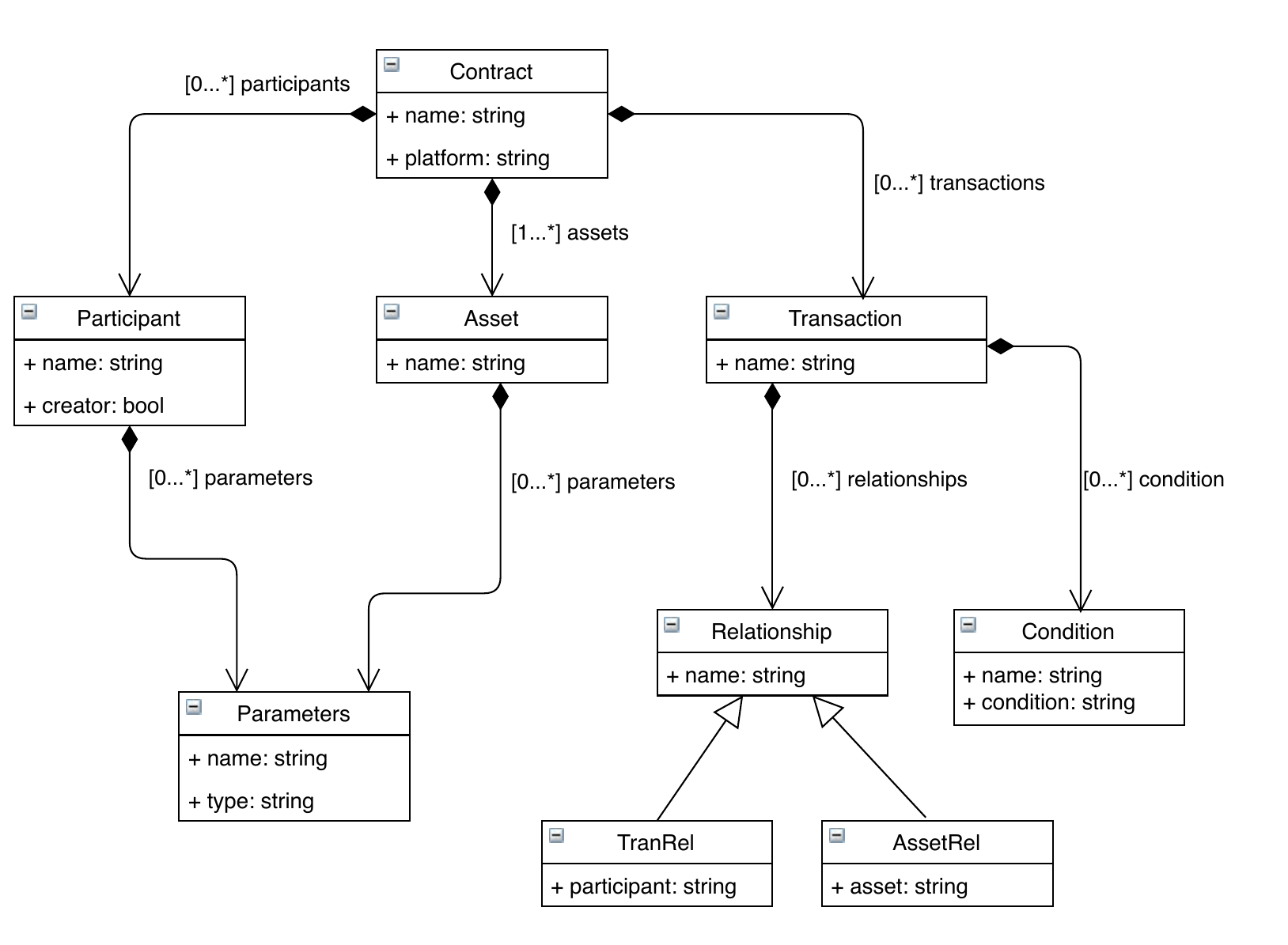}}
\caption{Intention DSL}
\label{DSL}
\end{figure*}
\begin{itemize}
\item Contract: where the user specifies a name for the contract and a platform. We support three blockchain platforms, which are Azure, Hyperledger Fabric, and Ethereum. 
\item Participant: the user can specify multiple participants, where each participant has a name (or identifier), and a list of parameters that describe the participant.
\item Asset: a tangible or intangible value that the user can specify. Any object of value in the real world may be represented as an asset.
\item Transaction: a user will specify a transaction, which is a function that can modify the values of the attributes of a participant or an asset.
\item Relationship: a user will define if the transaction has a relationship with a participant (TranRel) or with an asset (AssetRel).
\item Condition: the end-user will specify if there any access condition on a defined transaction.
\end{itemize}

After the chatbot framework detects the modeler intent, we map it to the structure of the model specification and create an instance of the DSL model. 

Table \ref{tab1} demonstrates the mapping between the vehicle auction use case and the main classes of the intention DSL. 

\begin{table}
\caption{Mapping the Intention DSL classes with Vehicle Auction Use-Case}
\scriptsize
\begin{center}
\begin{tabular}{|c|l|}
\hline
DSL Intention Class & Vehicle Auction Use-Case \\\hline
Contract & \begin{tabular}[c]{@{}l@{}}name: Vehicle Auction\\ platform: Ethereum\end{tabular} \\ \hline
Asset & name: Vehicle \\ \hline
\multirow{2}{*}{Participants} & \begin{tabular}[c]{@{}l@{}}Participant 1:\\ name: Owner\\ creator : True\end{tabular} \\ \cline{2-2} 
 & \begin{tabular}[c]{@{}l@{}}Participant 2:\\ name: Bidder\\ creator : False\end{tabular} \\ \hline
\multirow{2}{*}{Transactions} & \begin{tabular}[c]{@{}l@{}}Transaction 1:\\ name: Place-bid\end{tabular} \\ \cline{2-2} 
 & \begin{tabular}[c]{@{}l@{}}Transaction 2:\\ name: Withdraw\end{tabular} \\ \hline
\multirow{2}{*}{Relationship} & \begin{tabular}[c]{@{}l@{}}TranRel for Place-bid transaction: \\ participant : Bidder\end{tabular} \\ \cline{2-2} 
 & \begin{tabular}[c]{@{}l@{}}TranRel for Withdraw transaction: \\ participant : Owner\end{tabular} \\ \hline
\end{tabular}
\label{tab1}
\end{center}
\end{table}

The DSL model instance is validated against the iContractML model. From the validated model we apply a model to text transformation using Xtend to generate the smart contract code. The transformation template used in Xtend is described in \cite{hamdaqa2020icontractml}. 

\section{Conclusion}\label{conclusion}
In this paper, we have investigated how chatbot is utilized to facilitate the usage of MDE in code development. We have introduced iContractBot, a chatbot framework for smart contract development, and we have integrated it with iContractML, a DSML for developing smart contracts. This is achieved by building a DSL for the captured user intent and then generating an instance of the iContractML model based on the DSL by applying Model-to-Model transformation. A vehicle auction smart contract was developed using iContractBot as a case study to demonstrate the framework.

For future direction, we are planning to conduct an empirical study from multiple perspectives (user, contract language, etc.) that compares the two modalities, that is the graphical and the conversational interface. 

\section*{Data Availability}
The iContractBot project scripts are openly available at iContractBot repository \footnote{https://zenodo.org/record/4595966}.

\bibliographystyle{plain}
\bibliography{main.bib}

\vspace{12pt}

\end{document}